\documentclass[reprint,
 amsmath, amssymb,
 aps, floatfix, showkeys, pra
]{revtex4-2}

\usepackage{graphicx}
\usepackage{dcolumn}
\usepackage{bm}
\usepackage[colorlinks, linkcolor = magenta, urlcolor = blue, citecolor = blue]{hyperref}
\usepackage{braket}
\usepackage{physics}
\usepackage{amsmath}
\usepackage{array}
\usepackage{float}
\usepackage{multirow}

\begin{document}

\title{Interrelation of Non-Classicality, Entropy, Irreversibility and Work extraction in Open Quantum Systems}
\author{Jai Lalita}
\email{jai.1@iitj.ac.in}
\author{Subhashish Banerjee}
\email{subhashish@iitj.ac.in }
\affiliation{Indian Institute of Technology, Jodhpur-342030, India}

\date{\today}

\begin{abstract}
The interplay of non-classical volume, von Neumann entropy, entropy production, and ergotropy is investigated in various open quantum systems. Two categories of open quantum system models are utilized: spin-spin and spin-boson interaction models. The spin-spin interaction models include the quantum collision and central spin models. On the other hand, the spin-boson interaction models consist of non-Markovian amplitude damping channel, Markovian generalized amplitude damping channel, and the Jaynes-Cummings model. Across these various open quantum systems, universal interrelations emerge, where the non-classical volume shows contrasting evolution with entropy, and entropy production contrasts with ergotropy. The initial state of the reservoir in these open quantum systems is shown to have an impact on these interrelations. These findings establish an interesting link between quantum information and the thermodynamics of open quantum systems.
\end{abstract}

\keywords{Non-classicality, irreversibility, work extraction, open quantum systems }

\maketitle
\section{\label{sec:intro}Introduction\protect}
The study of open quantum systems forms a cornerstone of modern quantum physics, as no physical system can be perfectly isolated from its surrounding environment. The inevitable system–bath interaction gives rise to decoherence, dissipation, and irreversibility phenomena that play a pivotal role in determining the evolution of quantum states and the exchange of energy and information with the environment~\cite{Breuer2007, nielsen2010quantum, weiss2012quantum, Banerjee2018}. The dynamics of such open systems are typically described either by master equations or through completely positive and trace-preserving (CPTP) maps, both of which ensure the physical consistency of time evolution. The master equation formalism provides a continuous-time differential description of the system's reduced dynamics, while CPTP maps offer a discrete or operational representation that preserves quantum probabilities and positivity. Together, they form the mathematical foundation for analyzing how quantum systems evolve under environmental influences~\cite{Breuer1999Stochastic, Breuer_2012_foundations, Srikanth2008Squeezed, czerwinski2022dynamics, plenio2008dephasing, rivas2014quantum, utagi2020temporal, banerjee2017characterization}.

In this work, we investigate two broad classes of open quantum system models, namely, spin–spin and spin–boson interaction models, to encompass a wide range of physically relevant and experimentally realizable scenarios. The spin–spin interaction category includes the quantum collision model~\cite{ThermalizingQuantumMachines_2002, ziman2005description, Rybár_2012, Ciccarello2013Collision-model, McCloskey2014Non-Markovianity, Campbell2018Systemenvironment, csenyacsa2022entropy, lalita2025non_classicality} and the central spin model~\cite{NV2000Theory, Breuer2004Non_Markovian, He2019Exact, Mukhopadhyay2017Dynamics, Tiwari2022Dynamics, tiwari2024strong}, both of which are instrumental in describing system–environment coupling in finite-dimensional Hilbert spaces. The spin–boson interaction category, on the other hand, includes models such as the non-Markovian amplitude damping channel~\cite{Garraway1997Decay, nielsen2010quantum, Breuer_2012_foundations}, the Markovian generalized amplitude damping channel~\cite{Srikanth2008Squeezed, omkar2013dissipative}, and the Jaynes–Cummings model~\cite{Larson2021TheJaynes–Cummings, Jaynes1963Comparison, Garraway1997Decay}, which collectively capture memory effects and energy exchange between discrete quantum systems and bosonic reservoirs. Here, the collision models hold particular significance among spin–spin interaction frameworks due to their ability to simulate open-system dynamics through repeated and controllable system-environment interactions~\cite {Ciccarello2013Collision-model, McCloskey2014Non-Markovianity, lalita2025non_classicality}. Additionally, they can be demonstrated as numerically exact techniques~\cite{lacroix2025making}. Moreover, the collision model possesses an intrinsic thermalization capability, enabling the system to asymptotically reach a thermal steady state determined by the statistical properties of the ancillas, thereby providing a microscopic and operational perspective on quantum thermalization~\cite{arisoy2019thermalization, lalita2025non_classicality, BANERJEE2023Thermalization}.

An essential concept underlying the analysis of the above-specified models is non-classicality~\cite{Wigner1932Quantum, Sudarshan1963Equivalence, Thapliyal2015Quasiprobability}, which embodies purely quantum features such as superposition, coherence, and entanglement that have no classical counterpart. Non-classicality serves as a fundamental resource in quantum technologies~\cite{dowling2003quantum}, enabling quantum advantage in quantum computation~\cite{Horodecki2009QuantumEntanglement, Chandrashekar2007Symmetries}, secure communication~\cite{Ekert1991Quantumcryptography, Scarani2009The_security}, and precision measurement~\cite{Vittorio2004Quantum_Enhanced, Pezz2018Quantummetrology}. In the context of open quantum systems, it plays a dual role: while environmental interactions typically degrade non-classical correlations through decoherence, the persistence or regeneration of non-classical features, especially in non-Markovian regimes, offers deep insights into reversibility, memory effects, and energy exchange processes. Quantitative witness such as the non-classical volume~\cite{Anatole2004Negativity} are therefore indispensable for connecting the microscopic quantum behavior of systems to their macroscopic thermodynamic consequences.

A comprehensive characterization of open-system dynamics aims to bridge the informational and thermodynamic viewpoints. Four quantities, non-classical volume, von Neumann entropy, entropy production, and ergotropy, jointly provide complementary insights into this connection. The non-classical volume quantifies coherence and superposition~\cite{Anatole2004Negativity}; the von Neumann entropy characterizes statistical uncertainty and information loss~\cite{nielsen2010quantum}; entropy production captures the degree of irreversibility~\cite{Esposito2010Threefaces}; and ergotropy represents the extractable work through cyclic unitary operations~\cite{AEAllahverdyan_2004Maximalwork}. Together, these quantities elucidate how quantum resources transform into thermodynamic quantities. Recent investigations in quantum information, quantum thermodynamics, and quantum batteries have further highlighted the intricate connections between them ~\cite{Manfredi2000Entropy, Perarnau2015Extractable, Francica2020QuantumCoherence, Medina2025Anomalous, pathania2025quantum}. 

Building upon these insights, the present study systematically explores the interplay among non-classicality, von Neumann entropy, entropy production, and ergotropy across both spin–spin and spin–boson interaction models. By comparing the collision and central spin models, the non-Markovian amplitude damping and Markovian generalized amplitude damping channels, and the Jaynes–Cummings model, this work aims to uncover universal correspondences linking non-classicality, irreversibility, and extractable work. Such an investigation contributes toward a deeper understanding of how environmental structure, memory effects, and system–bath coupling jointly determine open quantum systems' informational and thermodynamic evolution.


The paper is organized as follows. Section~\ref{Preliminaries} introduces the four key quantities, i.e., non-classical volume, von Neumann entropy, entropy production, and ergotropy. Section~\ref{spin-spin} analyzes spin–spin interaction models, namely the quantum collision and central spin models. Section~\ref{spin-boson} extends the investigation to spin–boson interactions, including the non-Markovian and generalized amplitude-damping channels and the Jaynes–Cummings model. Section~\ref{conclusion} concludes with a discussion on the conceptual implications of these findings for the design of resource-efficient quantum thermodynamic devices.
\section{\label{Preliminaries} Preliminaries}
\subsection{Non-classical Volume}
The non-classical volume is a quantitative witness of non-classicality, initially introduced in the context of phase space quasi-probability distributions~\cite{Anatole2004Negativity}. It is defined using the Wigner function, $W(\theta, \phi)$, representing the quantum state in continuous phase space~\cite{Wigner1932Quantum, Thapliyal2015Quasiprobability}. Quantum states with non-classical characteristics, such as superposition and entanglement, can create regions of negative values of the Wigner functions, whereas classical states have a non-negative Wigner function. 

The non-classical volume, denoted by ${\delta}$, is computed as follows~\cite{Thapliyal2015Quasiprobability, lalita2025non_classicality},
\begin{equation}
    {\delta} = \int |W({\theta}, {\phi})| \sin{\theta} d{\theta} d{\phi} - 1.
    \label{NV_formula}
\end{equation}
This witness vanishes for states that can be described by classical probability distributions (e.g., coherent states) and increases with the degree of non-classicality. Importantly, ${\delta}$ is basis-independent, making it suitable for comparing different systems or studying the dynamical degradation of non-classicality under noise. It has found applications in quantifying decoherence in optical fields and qubits evolving under various open-system models~\cite{Thapliyal2016tomograms}.

\subsection{\label{Entropy}Von Neumann Entropy}
A fundamental concept in quantum information, the von Neumann entropy extends the classical Shannon entropy to quantum systems. For a quantum state represented by a density operator $\rho$, the von Neumann entropy is defined as~\cite{nielsen2010quantum} 
\begin{equation}
    S(\rho) = - \mathrm{Tr}(\rho \log \rho).
    \label{von_neumann_entropy}
\end{equation}
It captures the degree of uncertainty or mixedness associated with the state. For this reason, pure states have zero entropy, while maximally mixed states attain the highest entropy allowed by the system's Hilbert space dimension. 

Unlike classical entropy, the von Neumann entropy accounts for quantum superposition and entanglement, as a key tool for characterizing correlations between subsystems. In bipartite systems, it plays a central role in quantifying entanglement through the entropy of reduced density matrices~\cite{Horodecki2009quantum_ent}. Moreover, its concavity and invariance under unitary transformations underline its consistency as an information-theoretic measure. The von Neumann entropy not only bridges quantum physics and information theory but also appears in the thermodynamics of open quantum systems~\cite{Breuer2007}, and black hole physics~\cite{Bekenstein1973Blackholes, Jha2025probing}, making it indispensable for understanding both the informational and physical aspects of quantum mechanics.

\subsection{\label{Entropy production} Entropy production}
The quantum formulation of entropy production begins with the joint unitary evolution of a system $S$ and its environment $E$, initially in states $\rho_S$ and $\rho_E$. Their composite state after interaction through a global unitary operator $U$ is
\begin{equation}
    \rho'_{SE} = U(\rho_S \otimes \rho_E)U^\dagger.
\end{equation}
The system's reduced state follows from tracing out the environment, $\rho'_S = \mathcal{E}(\rho_S) = \mathrm{tr}_E\{\rho'_{SE}\}$, and it is precisely this partial trace that introduces irreversibility, since discarding inaccessible environmental degrees of freedom and system-environment correlations prevents exact reversibility. Entropy production in this framework is quantified as~\cite{Esposito2010Threefaces, Landi2021Irreversibleentropy}
\begin{equation}
    \Sigma = I^{\rho'_{SE}}(S: E) + S(\rho'_E \| \rho_E),
\end{equation}
where the first term is the mutual information created between the system and its environment,
$I^{\rho'_{SE}}(S: E) = S(\rho'_S) + S(\rho'_E) - S(\rho'_{SE})$, and the second term is the relative entropy between the final and initial environment states, $S(\rho \| \sigma) = \Tr[\rho \ln \rho - \rho \ln \sigma]$. These contributions together can be written more compactly as
\begin{equation}
    \Sigma = S(\rho'_{SE} \| \rho'_S \otimes \rho_E),
    \label{entropy_production_final}
\end{equation}
where $\rho'_{SE}$ and $\rho'_S$ represent the evolved states of the composite system-environment and the system, respectively. Furthermore, $\rho_E$ is the initial state of the environment fed to the system during evolution at each step. Altogether, these relations establish entropy production as a non-negative measure of lost information, arising from correlations with the environment and irretrievable changes in it.

\subsection{\label{Ergotropy} Ergotropy}
Ergotropy quantifies the maximum extractable work from a quantum state under unitary operations, without changing its entropy~\cite{AEAllahverdyan_2004Maximalwork}. For open quantum systems, ergotropy is determined by using the evolved state ($\rho'_s$) of the system as the input for the computation. This enables us to ascertain the limitations on work extraction through unitary transformations, including environmental influences~\cite{cakmak2020Ergotropy}. The ergotropy of a $d$-dimensional quantum system is given by~\cite{AEAllahverdyan_2004Maximalwork, Tiwari2023Impact}
\begin{equation}
    \mathcal{W}(\rho'_s) = \text{Tr}[\rho'_s H_s] - \min_{U} \text{Tr}[U \rho'_s U^\dagger Hs],
\end{equation} 
where the minimization identifies the passive state of $\rho'_s$. The passive state is obtained by rearranging the eigenvalues of $\rho'_s$ in decreasing order and aligning them with the eigenvectors corresponding to the increasing eigenvalues of $H_s$. This ensures no further work can be extracted through unitary operations.

For a single-qubit system, the Hamiltonian is generally expressed as $H_s = \frac{\omega_0}{2} \sigma_z$, where $\omega_0$ is the qubit frequency and $\sigma_z$ the Pauli operator. The mean energy of the system at time $t$ is $E(t) = \text{Tr}[\rho'_s H_s]$, while the corresponding passive energy is defined as
$E_{\text{pas}}(t) = \sum_i r_i \epsilon_i$, with $r_i$ being the ordered eigenvalues of $\rho'_s$, largest to smallest, \textit{viz.} $r_1 \geq r_2 \geq r_3 ...\geq r_d$ and $\epsilon_i$ the energy eigenvalues of $H_s$ smallest to largest, \textit{i.e.}, $\epsilon_1 \leq \epsilon_2 \leq \epsilon_3 ...\leq \epsilon_d$.
Finally, the ergotropy for a single-qubit state takes the explicit form~\cite{tiwari2024strong, Andolina_OQS_battery}
\begin{equation}
    \mathcal{W}(t) = \frac{\omega_0}{2} \left(r'_{z} + \sqrt{(r_{x}^{'2} + r_{y}^{'2} + r_{z}^{'2})} \right),
    \label{ergotropy_formula}
\end{equation}
where $r'_x, r'_y, r'_z$ are the Bloch vector components of $\rho'_s$.

\section{\label{spin-spin}Spin-Spin interaction Models}
This section explores the above-specified quantities under the spin-spin interaction models, particularly the quantum collision and central spin models. 
\subsection{\label{Collision model} Collision model}
\begin{figure}
    \centering
    \includegraphics[width=1\linewidth]{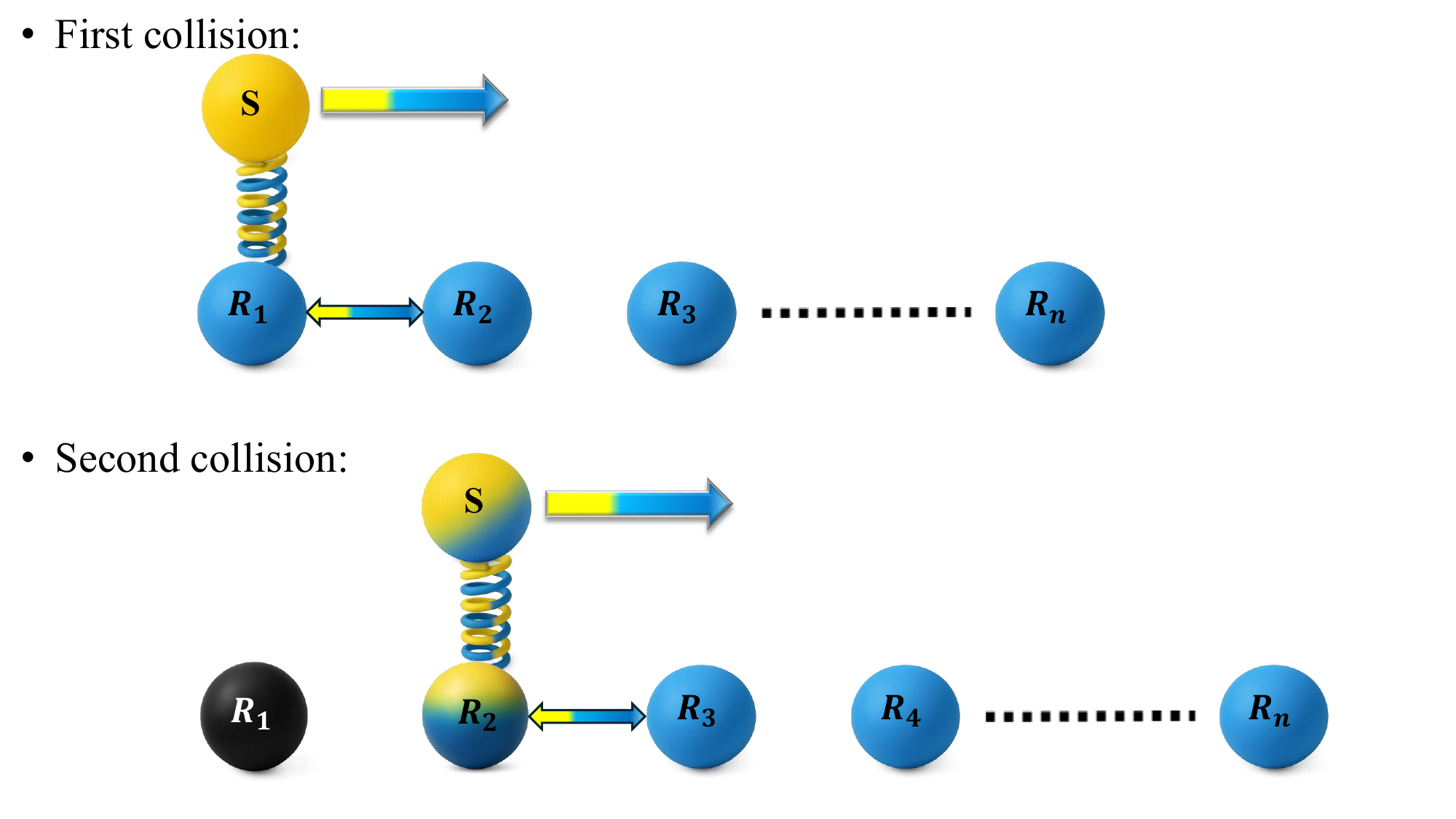}
    \caption{This diagram illustrates a single-qubit collision model in which the system qubit, denoted as $S$, interacts with a sequence of ancilla qubits $R_n$. The spiral lines represent a Heisenberg-type interaction between the system qubit $S$ and the ancillae. In contrast, the double-sided arrowed straight lines indicate a partial-swap interaction between successive ancilla qubits.}
    \label{Single_qubit_collision_model}
\end{figure}
The collision model provides a transparent and versatile way to describe the reduced dynamics of an open quantum system. The idea is to model the reservoir not as a single large environment but as a collection of small, identical subunits (called ancillas) interacting with the system $S$ one at a time. In this framework, the reservoir is represented by a sequence of ancillas $\{R_1, R_2, R_3, \dots \}$, each interacting with the system $S$ for a finite time interval $\tau$~\cite{Ciccarello2013Collision-model, McCloskey2014Non-Markovianity, Campbell2018Systemenvironment,  Ciccarello2022Quantumcollisionmodel}, as shown in Fig.~\ref{Single_qubit_collision_model}. Significantly, every ancilla interacts only once with the system before being discarded, and the next ancilla in the sequence takes its place as illustrated by the black ball ($R_1$) in the second collision in Fig.~\ref{Single_qubit_collision_model}. The total Hamiltonian of the combined setup is written as
\begin{equation}
    H_T^{CM} = H_S + H_{R_n} + H_{SR_n},
\end{equation}
where $H_S = \frac{{\hbar\omega}_{s}}{2}\sigma_z$ is the Hamiltonian of the system of interest, $H_{R_n} = \frac{\hbar\omega_{R_n}}{2}\sigma_z$ represents each reservoir's subunit Hamiltonian, and the Heisenberg interaction $H_{SR_n} = g_{SR}(\sigma_x^{S}\sigma_x^{R_n} + \sigma_y^{S}\sigma_y^{R_n})$,  accounts for the system–reservoir interaction. Further, $\omega_s$, $\omega_{R_n}$, and $g_{SR}$ denote the system frequency, ancilla frequency, and system-reservoir coupling strength, respectively. Now, to study the evolution of system $S$, let us consider the $n^{th}$ collision between the system and the ancilla $R_n$. The joint unitary evolution operator for this process is given by
\begin{equation}
    U_{SR_n} = \exp\left[-i~\left(H_T^{CM}\right)\tau \right],
\end{equation}
where $\tau$ is the collision (interaction) time. If $\rho_S^n$ and $\rho_{R_n}$, respectively, denote the states of the system and the ancilla $R_n$ before their interaction, then the global state of the pair after the collision is $\rho_{SR_n}' = U_{SR_n} \, \big(\rho_S^n \otimes \rho_{R_n}\big) \, U_{SR_n}^\dagger$. The state of the system after this step is obtained by tracing out the ancilla as
\begin{equation}
    \rho_S^{n+1} = \mathrm{Tr}_{R_n}\!\left[\rho_{SR_n}'\right].
\end{equation}
This construction makes the system's dynamics iterative; each new step depends on the system's state from the previous step and the fresh ancilla it encounters.
In our case, we assume that all ancillas are prepared identically in the same thermal state, i.e., 
$\rho_{R_n}(0) = e^{-\beta_{R_n}H_{R_n}}/\Tr[e^{-\beta_{R_n}H_{R_n}}]$, and $\beta_{R_n} = 1/k_BT_{R_n}$ is the inverse temperatures of the environment qubit. Further, we set $\hbar = k_B = 1$ throughout the paper. In its simplest version, once an ancilla has interacted with the system, it is discarded and plays no further role. The environment, therefore, has no memory, and the system dynamics are purely Markovian. Each collision is independent, and the reduced evolution is described by a sequence of completely positive trace-preserving (CPTP) maps. However, one of the strengths of the collision model is that it can also be extended to describe non-Markovian dynamics by introducing correlations into the environment~\cite{McCloskey2014Non-Markovianity, ThermalizingQuantumMachines_2002}. This is achieved by allowing inter-ancilla interactions between successive system–ancilla collisions.
Here, we consider the scheme where the system still interacts with the ancillas in the same sequential manner described above. However, this ancilla is not immediately discarded after the system has interacted with ancilla $R_n$. Instead, it can undergo an additional unitary interaction with the next ancilla $R_{n+1}$ before $R_{n+1}$ collides with the system, as illustrated in Fig.~\ref{Single_qubit_collision_model}. Let the unitary interaction between the successive reservoir subunits be governed by a partial swap operation described as~\cite{nielsen2010quantum, loss1998quantum}
\begin{equation}
    \begin{aligned}
        U_{R_n, R_{n+1}} = \cos(\Theta) I - i\sin(\Theta)H_{swap},   
    \end{aligned}
    \label{p_swap_unitary}
\end{equation}
where $\Theta \in [0, \frac{\pi}{2}]$, and $H_{swap} = \frac{1}{2}(\Vec{\sigma}{^{R_n}}.~\Vec{\sigma}{^{R_{n+1}}} + I_4)$. This modification introduces correlations between successive ancillas. The key effect is that some information about the system, which has been imprinted on $R_n$ during its collision, can be passed on to $R_{n+1}$. When $R_{n+1}$ subsequently interacts with the system, this stored information can flow back, leading to non-Markovian dynamics.

Now, we investigate the dynamics of non-classical volume $\delta$, Eq.~\eqref{NV_formula}, von-Neumann entropy $S$, Eq.~\eqref{von_neumann_entropy}, entropy production $\Sigma$, Eq.~\eqref{entropy_production_final}, and ergotropy $\mathcal{W}$, Eq.~\eqref{ergotropy_formula}, for this model. The initial state of the system evolving under the non-Markovian collision model is taken to be the maximally negative quantum state, $NS_1$ state. 
The single-qubit $NS_1$ state is the eigenstate corresponding to the most negative eigenvalue of the phase space point operator according to the single-qubit discrete Wigner function formalism, detailed in~\cite{lalita2023harnessing, Lalita_2024ProtectingQC, lalita2025realizingnegativequantumstates}. The explicit form of the single-qubit $NS_1$ state, used in this work, in the Bloch vector form, is given by
\begin{equation}
    \rho_{NS_1} = \tfrac{1}{2}( I_2 + a_1\sigma_x + a_2\sigma_y + a_3\sigma_z),
\end{equation}
where the Bloch vector components $a_1$, $a_2$, $a_3$ take the following values $a_1 = 0.50$, $a_2 = 0.56$, $a_3 = -0.66$. 

\begin{figure}
    \centering
    \includegraphics[width=1\linewidth]{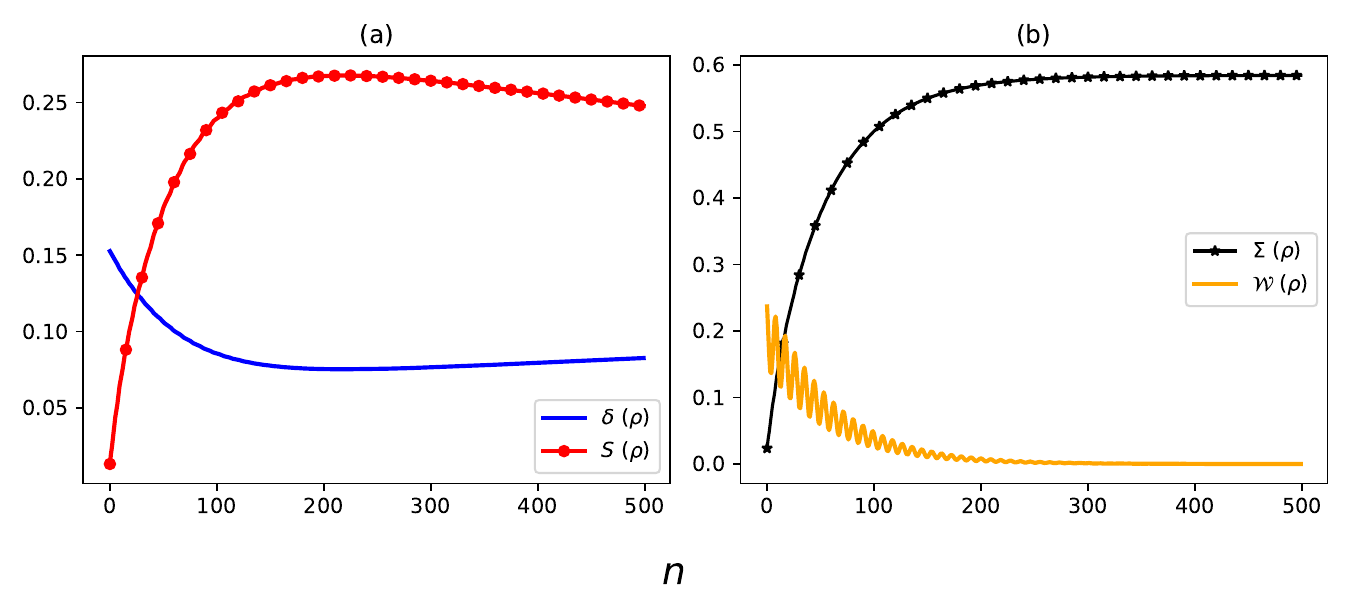}
    \caption{Variation of non-classical volume ($\delta$), von-Neumann entropy ($S$) in subplot $(a)$, and entropy production ($\Sigma$), ergotropy ($\mathcal{W}$) in subplot $(b)$ with the number of collisions of the single-qubit maximally negative quantum state using a non-Markovian collision model. The parameters are: $\omega_s = 1.5$, $\omega_R = 1$, $\beta = 50$, $g_{SR} = 0.5$, $\Theta = 0.98\frac{\pi}{2}$ and $\tau = 0.5$.}
    \label{non_Markovian_collision_model}
\end{figure}

Figure~\ref{non_Markovian_collision_model} illustrates the variation of non-classical volume, von-Neumann entropy, entropy production, and ergotropy with the number of collisions $n$ of the single-qubit non-Markovian collision model. An intriguing contrasting behavior is observed between the non-classical volume $\delta$ and von Neumann entropy $S$, as shown in Fig. \ref{non_Markovian_collision_model}(a). Specifically, a decrease in non-classicality corresponds to an increase in the system’s randomness. Figure \ref{non_Markovian_collision_model}(b) highlights a similar opposite relation between entropy production and ergotropy, demonstrating that reduced irreversibility facilitates greater work extraction. Moreover, a closer examination of Figs. \ref{non_Markovian_collision_model}($a$) and ($b$) reveals that higher von Neumann entropy is accompanied by increased entropy production, while a decline in non-classical volume is associated with a corresponding reduction in ergotropy.  

\subsection{\label{Central spin model} Central spin model}
The central spin model is a fundamental framework in quantum many-body physics. It is often described as a single two-level system (the central spin) interacting with a bath of surrounding spins, arranged symmetrically around it~\cite{NV2000Theory, Tiwari2022Dynamics, tiwari2024strong, Breuer2004Non_Markovian}. For this model, the Hamiltonian of the composite system is given by
\begin{equation}
    H = H_S + H_B + V,
\end{equation}
with individual contributions (for $\hbar = 1$)
\begin{equation}
    H = \frac{\omega_0}{2}\sigma_z^0 + \frac{\omega}{N}J_z + \frac{\boldsymbol{\epsilon}}{\sqrt{N}}\left(\sigma_x^0 J_x + \sigma_y^0 J_y\right).
\end{equation}
Here, $\omega_0$ is the transition frequency of the central spin, $\omega/N$ is the scaled frequency of the collective bath and $N$ is the number of bath spins, $\boldsymbol{\epsilon}$ is the system–bath coupling strength, $\sigma_\alpha^0$ $(\alpha = x,y,z)$ are Pauli operators of the central spin, and $J_\alpha = \frac{1}{2}\sum_{i=1}^N \sigma_\alpha^{(i)}$ are collective angular momentum operators of the bath spins. The global unitary evolution of the joint system–bath state for an initial separable condition $\rho_{SB}(0) = \rho_S(0) \otimes \rho_B(0)$ is given by
\begin{equation}
    \rho_{SB}(t) = e^{-iHt}\,\rho_{SB}(0)\,e^{iHt}.
\end{equation}
The reduced dynamics of the central spin is then obtained by tracing out the bath degrees of freedom as
\begin{equation}
    \rho'_S = \text{Tr}_B\left[e^{-iHt}\,\rho_{SB}(0)\,e^{iHt}\right].
\end{equation}
The bath spins are initially considered to be in a thermal state (Gibbs state), obtained using the spectral decomposition of the bath Hamiltonian
$H_B = \frac{\omega}{N} J_z = \frac{\omega}{2}\sum_{n=0}^N \left(1 - \frac{2n}{N}\right)|n\rangle\langle n|$ as
\begin{equation}
    \rho_B(0) = \frac{e^{-\beta H_B}}{Z} = \frac{1}{Z}\sum_{n=0}^N e^{-\frac{\beta \omega}{2}\left(1 - \frac{2n}{N}\right)}|n\rangle\langle n|,
\end{equation}
where $Z = \sum_{n=0}^N e^{-\frac{\beta \omega}{2}\left(1 - \frac{2n}{N}\right)}$ with inverse temperature $\beta = 1/k_BT$ and $\ket{n}$ is the standard computational basis.
This formulation sets the stage for deriving the exact dynamics of the central spin. The time-dependent reduced state $\rho'_S$ can be explicitly calculated by diagonalizing the total Hamiltonian $H$ numerically. This exact solvability makes the central spin model a powerful tool for probing strong-coupling and non-Markovian quantum thermodynamics. 

\begin{figure}
    \centering
    \includegraphics[width=1\linewidth]{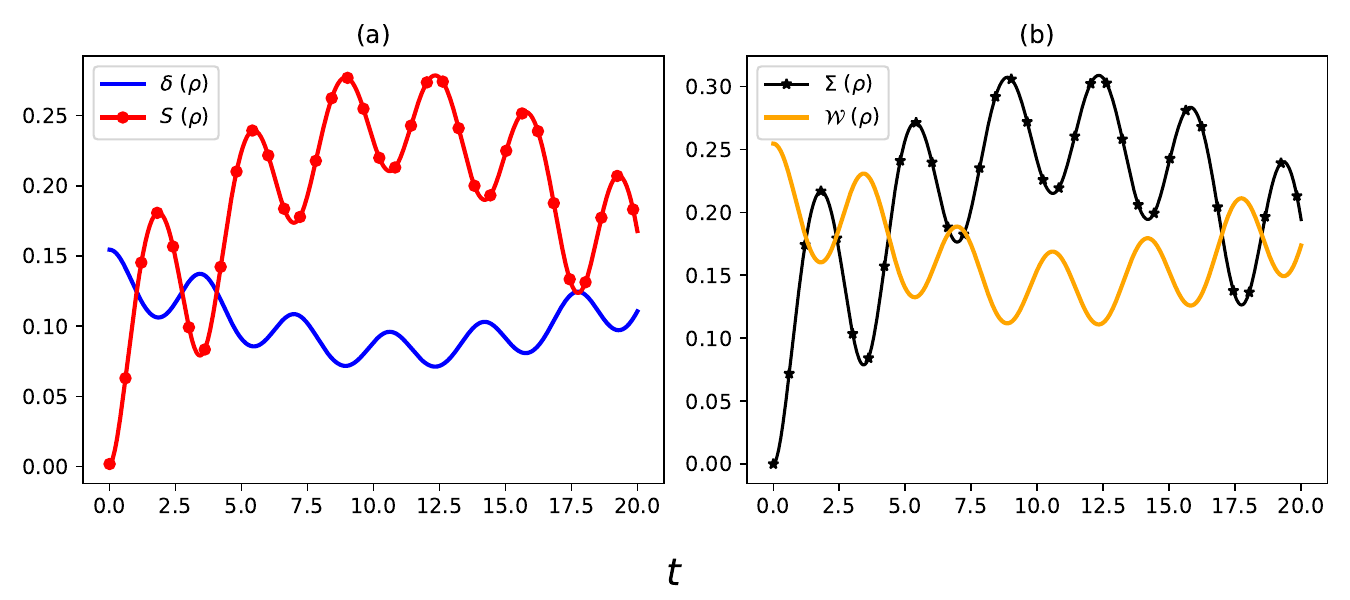}
    \caption{Variation of non-classical volume ($\delta$), von-Neumann entropy ($S$) in subplot $(a)$, and entropy production ($\Sigma$), ergotropy ($\mathcal{W}$) in subplot $(b)$ with time using the central spin model of the single-qubit maximally negative quantum state. The parameters are: $\omega_0 = 1.5$, $\omega = 1$, $\beta = 100$, $\boldsymbol{\epsilon} = 0.5$ and $N = 50$.}
    \label{Central_spin_model}
\end{figure}

The dynamics of the non-classical volume and above-specified thermodynamic quantities for the single-qubit $NS_1$ state using the central spin model are studied using Eqs. \eqref{NV_formula}, \eqref{von_neumann_entropy}, \eqref{entropy_production_final}, and \eqref{ergotropy_formula}. Within this spin–spin interaction model, we also observe a contrasting relationship between $\delta$ and $S$, as well as between $\Sigma$ and $\mathcal{W}$, as shown in Fig.~\ref{Central_spin_model}, which mirrors the correspondence found in the collision model. Moreover, the mutual interdependence among $S$ and $\Sigma$, $\delta$, and $\mathcal{W}$ is preserved, highlighting the consistent thermodynamic structure underlying the system’s dynamics in a thermal spin environment.

\section{\label{spin-boson}Spin-boson interaction Models}
Next, we study the interrelation between non-classical volume and the above-specified quantum thermodynamic quantities for a spin interacting with the bosonic environment. In particular, we consider the non-Markovian amplitude damping (NMAD) channel, the generalized amplitude damping (GAD) channel, and the Jaynes-Cummings Model (JCM).

\subsection{\label{Garraway Model} Non-Markovian amplitude damping model}
This model studies the decay of a two-level atom interacting with a bosonic reservoir~\cite{Garraway1997Decay, Breuer_2012_foundations}. The total Hamiltonian governing the system bath setup is given by
\begin{equation}
    H = H_s \otimes I_B + I_s \otimes H_B + H_I,
\end{equation}
where $H_s$ is the system Hamiltonian, $H_B$ is the bath Hamiltonian, and $H_I$ describes their interaction. For the qubit system,
\begin{equation}
    H_s = \frac{\omega_0}{2} \sigma_z,
\end{equation}
where $\omega_0$ is the system qubit frequency. The environment is modeled as a collection of harmonic oscillators,
\begin{equation}
    H_B = \sum_k \omega_k a_k^\dagger a_k,
\end{equation}
where $a_k^\dagger$ and $a_k$ are bosonic creation and annihilation operators satisfying $[a_k,a_{k'}^\dagger] = \delta_{k,k'}$. The interaction Hamiltonian is given by
\begin{equation}
    H_I = \sum_k \left( g_k \sigma^+ \otimes a_k + g_k^* \sigma^- \otimes a_k^\dagger \right),
\end{equation}
where $g_k$ is the coupling constant and $\sigma^+ = |e\rangle\langle g|$ and $\sigma^- = |g\rangle\langle e|$ denote the atomic raising and lowering operators, with $\ket{g}$ and $\ket{e}$ being the ground and excited states, respectively. Here, the initial state of the bath is taken to be the vacuum state. The total number of excitations, for this model, is conserved, which in this case is one. The reduced dynamics of the system follow the following master equation~\cite{Breuer1999Stochastic}
\begin{equation}
    \begin{aligned}
       \frac{d}{dt}\rho_s(t) &= -\frac{i}{2} S(t)[\sigma^+\sigma^-,\rho_s(t)]\\
       &+ \gamma(t)\Big(\sigma^- \rho_s(t) \sigma^+ - \tfrac{1}{2}\{\sigma^+\sigma^-,\rho_s(t)\}\Big), 
    \end{aligned}
    \label{dynamics_eq}
\end{equation}
where $\gamma(t) = -2 \,Re\left(\frac{\dot{G}(t)}{G(t)}\right)$ and $S(t) = -2 \,Im\left(\frac{\dot{G}(t)}{G(t)}\right)$ are the decay rate and the time-dependent frequency shift, respectively. For a Lorentzian spectral density of the bath in resonance with the qubit frequency, the function $G(t)$ becomes
\begin{equation}
    G(t) = e^{-\lambda t/2}\!\left[ \cosh\!\left(\tfrac{lt}{2}\right) + \tfrac{\lambda}{l}\sinh\!\left(\tfrac{lt}{2}\right) \right],
\end{equation}
where $l = \sqrt{\lambda^2 - 2\gamma_0 \lambda}$, with $\gamma_0$ being the system-bath coupling strength and $\lambda$ being the spectral width of the bath. Further, $-2\frac{\dot{G}(t)}{G(t)} = 2\left(\frac{\gamma_0}{ \sqrt{1 - \tfrac{2\gamma_0}{\lambda}} \, \coth\!\left(\tfrac{1}{2}\lambda t\sqrt{1-\frac{2\gamma_0}{\lambda}}\right) + 1 }\right)$.
For $\lambda < 2\gamma_0$, the decay rate becomes negative in certain intervals, leading to non-Markovian amplitude damping (NMAD) evolution, while $\lambda > 2\gamma_0$ gives time-dependent Markovian dynamics, and $\lambda \gg \gamma_0$ reduces to the standard time-independent amplitude damping channel.
Moreover, the qubit state at time $t$, $\rho_s(t)$ using Eq.~\eqref{dynamics_eq} can be expressed in Bloch vector form as
\begin{equation}
    \rho_s(t) = \tfrac{1}{2} \begin{pmatrix} 1+z(t) & x(t)-iy(t) \\ x(t)+iy(t) & 1-z(t) \end{pmatrix},
    \label{rho_bloch_vector}
\end{equation}
with $x(t) = \text{Tr}[\sigma_x \rho(t)]$, $y(t) = \text{Tr}[\sigma_y \rho(t)]$, and $z(t) = \text{Tr}[\sigma_z \rho(t)]$. Solving the dynamics using Eq. \eqref{rho_bloch_vector} gives
\begin{equation}
   \begin{aligned}
       \rho_{00}(t) &= \big(1-|G(t)|^2\big)\rho_{11}(0) + \rho_{00}(0), \\
    \rho_{01}(t) &= \rho_{01}(0)\, G^*(t), \\
    \rho_{10}(t) &= \rho_{10}(0)\, G(t), \\
    \rho_{11}(t) &= \rho_{11}(0)|G(t)|^2,
    \label{rho_elements}
   \end{aligned}
\end{equation}
where $\rho_{ij}$'s are the elements of the density matrix $\rho(t)$. On comparing Eqs. \eqref{rho_bloch_vector} and \eqref{rho_elements}, the Bloch vector components are given as
\begin{equation}
   \begin{aligned}
    x(t) = 2\,Re[\rho_{10}(0) G(t)], \\
    y(t) = -2\,Im[\rho_{10}(0) G(t)], \\
    z(t) = 2\rho_{11}(0)|G(t)|^2 - 1.
   \end{aligned}
\end{equation}

\begin{figure}
    \centering
    \includegraphics[width=1\linewidth]{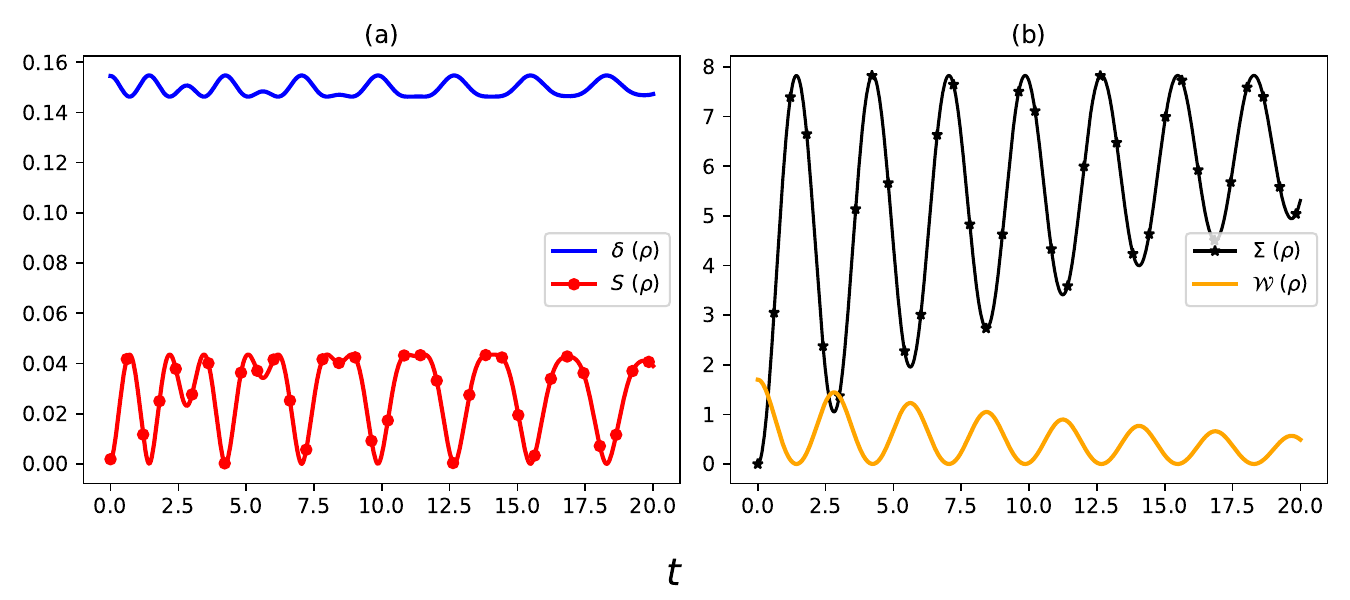}
    \caption{Variation of non-classical volume ($\delta$), von-Neumann entropy ($S$) in subplot $(a)$, and entropy production ($\Sigma$), ergotropy ($\mathcal{W}$) in subplot $(b)$ with time of the single-qubit maximally negative quantum state under the NMAD channel. The parameters are: $\omega_0 = 10$, $\lambda = 0.05$ and $\gamma_0 = 50$.}
    \label{NMAD_channel}
\end{figure}
Figure~\ref{NMAD_channel} shows the variation of the non-classical volume, von-Neumann entropy, entropy production, and ergotropy with time under the NMAD channel. The system's initial state is taken to be the $NS_1$ state. The NMAD channel satisfies the global fixed point condition, viz., $U[\rho_{s}^{*} \otimes \rho_E]U^{\dag} = \rho_{s}^{*} \otimes \rho_E$, where $\rho_{s}^{*}$ is the steady state of the system and $\rho_E$ is the initial state of the bath. For this condition, the formula for the entropy production becomes~\cite{Landi2021Irreversibleentropy},
\begin{equation}
    \Sigma = S(\rho_s \| \rho_{s}^{*}) - S(\rho'_s \| \rho_{s}^{*}),
    \label{entropy_production_rho_th}
\end{equation}
where $\rho_s$ and $\rho_s'$ are the initial and evolved states of the system.

From Fig.~\ref{NMAD_channel}(a), it can be observed that the variations of $\delta$ and $S$ exhibit a contradictory relationship. Similarly, the dynamics of $\Sigma$ and $\mathcal{W}$ also display contradictory behavior in the spin–boson interaction model, as shown in Fig.~\ref{NMAD_channel}(b), in close analogy with the spin–spin interaction model. However, a notable difference arises in their mutual dependence. In this case, the fluctuations of $S$ and $\Sigma$ no longer follow a similar trend, and the oscillations of $\delta$ and $\mathcal{W}$ are also not synchronized, as illustrated in Fig.~\ref{NMAD_channel}(a) and (b). A reason for this deviation could be the initial state of the bosonic bath, which is taken to be the ground state here. 

To take into account the impact of the thermal state as the initial state of the bosonic bath in the spin-boson interaction, we study the dynamics of the non-classical volume and the above-considered thermodynamic quantities under the generalized amplitude channel and the Jaynes-Cummings model.

\subsection{\label{GAD channel} Generalized amplitude damping channel}
The generalized amplitude damping (GAD) channel models a Markovian evolution due to a finite-temperature environment. It captures decay from the excited to the ground and thermal excitation from the ground to the excited states. The corresponding master equation is give by~\cite{Breuer2007, Srikanth2008Squeezed, omkar2013dissipative},
\begin{equation}
    \begin{aligned}
        \frac{d}{dt}\rho_s(t) &= -i[H_s,\rho_s(t)] \\
                            &+ \gamma(N^{th} + 1)\Big(\sigma^- \rho_s(t) \sigma^+ - \tfrac{1}{2}\{\sigma^+\sigma^-,\rho_s(t)\}\Big) \\
                            &+ \gamma N^{th} \Big(\sigma^+ \rho_s(t) \sigma^- - \tfrac{1}{2}\{\sigma^-\sigma^+,\rho_s(t)\}\Big),
    \label{GAD_dynamics_eq}
    \end{aligned}
\end{equation}
where $H_s = \frac{\omega_0}{2}\sigma_z$ is the system Hamiltonian, $\gamma$ is the dissipative constant, and $N^{th} = \frac{1}{e^{\beta \omega_0}- 1}$ is the mean thermal photon number of the bath at a finite temperature $T$. Also, the bosonic bath in this channel is taken to be the thermal state, i.e., $\rho_{B}(0) = e^{-\beta H_b}/\Tr[e^{-\beta H_b}]$, and $\beta = 1/k_BT$. The GAD channel also satisfies the global fixed point condition, and its entropy production is calculated using Eq.~\eqref{entropy_production_rho_th}.

\begin{figure}
    \centering
    \includegraphics[width=1\linewidth]{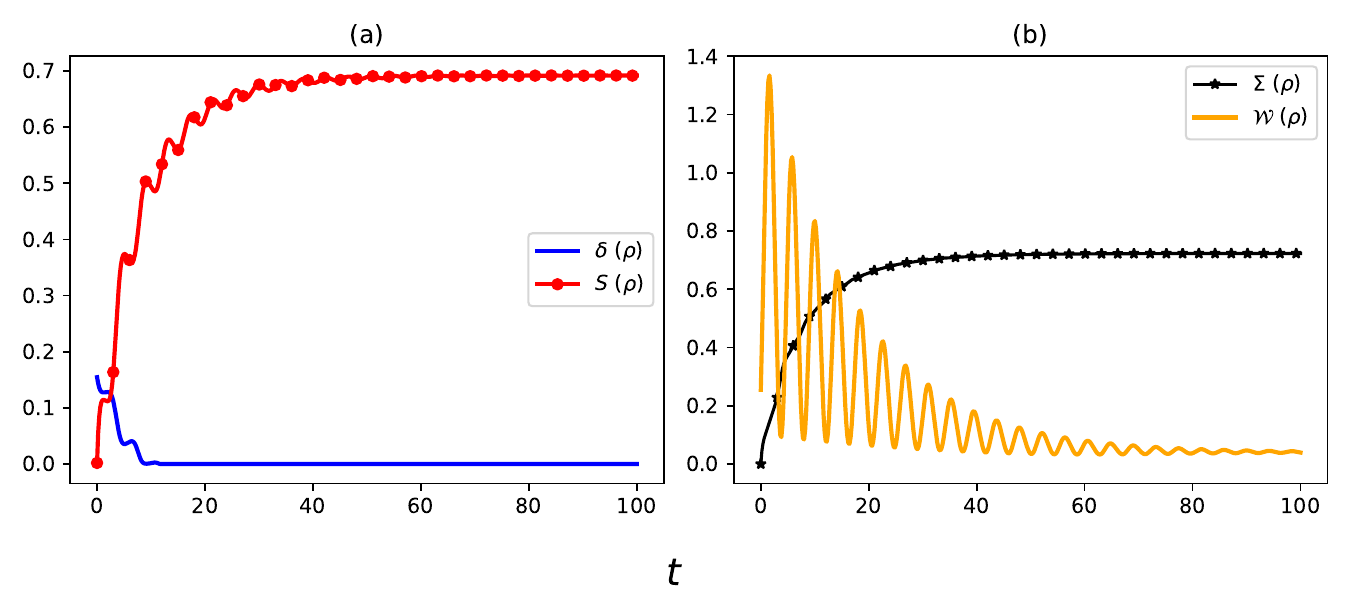}
    \caption{Variation of non-classical volume ($\delta$), von-Neumann entropy ($S$) in subplot $(a)$, and entropy production ($\Sigma$), ergotropy ($\mathcal{W}$) in subplot $(b)$ with time of the one-qubit maximally negative quantum state. The evolution is governed by the Markovian generalized amplitude damping master equation. The parameters are: $\omega_0 = 1.5$, $\beta = 1$, and $g = 0.05$.}
    \label{Markovian_GAD_channel}
\end{figure}

Figure~\ref{Markovian_GAD_channel} illustrates the variation of $\delta$, $S$, $\Sigma$, and $\mathcal{W}$ of the single-qubit $NS_1$ state with time under the Markovian GAD channel. We can observe from Fig.~\ref{Markovian_GAD_channel}(a) that $\delta$ and $S$ behave in an opposite manner. The $\Sigma$ and $\mathcal{W}$ also exhibit contradictory behavior as illustrated by Fig.~\ref{Markovian_GAD_channel}(b). Moreover, the rise and fall patterns of $\delta$ and $\mathcal{W}$ and those of $S$ and $\Sigma$ remain in agreement with the trends observed in the collision and central spin models. This consistency arises because the bosonic reservoir is initialized in a thermal state in the GAD channel, which permits energy exchange between the system and the bath. 

\subsection{\label{Jaynes-Cummings model}Jaynes-Cummings model}
The Jaynes-Cummings model is a single-mode Garraway model, i.e., a qubit inside a single bosonic mode of frequency $\omega_c$~\cite{Jaynes1963Comparison, Garraway1997Decay, Larson2021TheJaynes–Cummings}. The Hamiltonian of the Jaynes-Cummings model is
\begin{equation}
    H_{\mathrm{JC}}=\tfrac{\omega_0}{2}\sigma_z + \omega_c a^\dagger a + g(\sigma^+ a + \sigma^- a^\dagger),
\end{equation}
where $\tfrac{\omega_0}{2}\sigma_z$, $\omega_c a^\dagger a$, and $g(\sigma^+ a + \sigma^{-} a^\dagger)$ are the system $H_s$, bath $H_b$, and interaction $H_{sb}$ Hamiltonians, respectively. Here, the initial state of the bosonic bath is taken to be the thermal state, i.e., $\rho_{B}(0) = e^{-\beta H_b}/\Tr[e^{-\beta H_b}]$, and $\beta = 1/k_BT$, where $T$ is the temperature of the bosonic bath.

\begin{figure}
    \centering
    \includegraphics[width=1\linewidth]{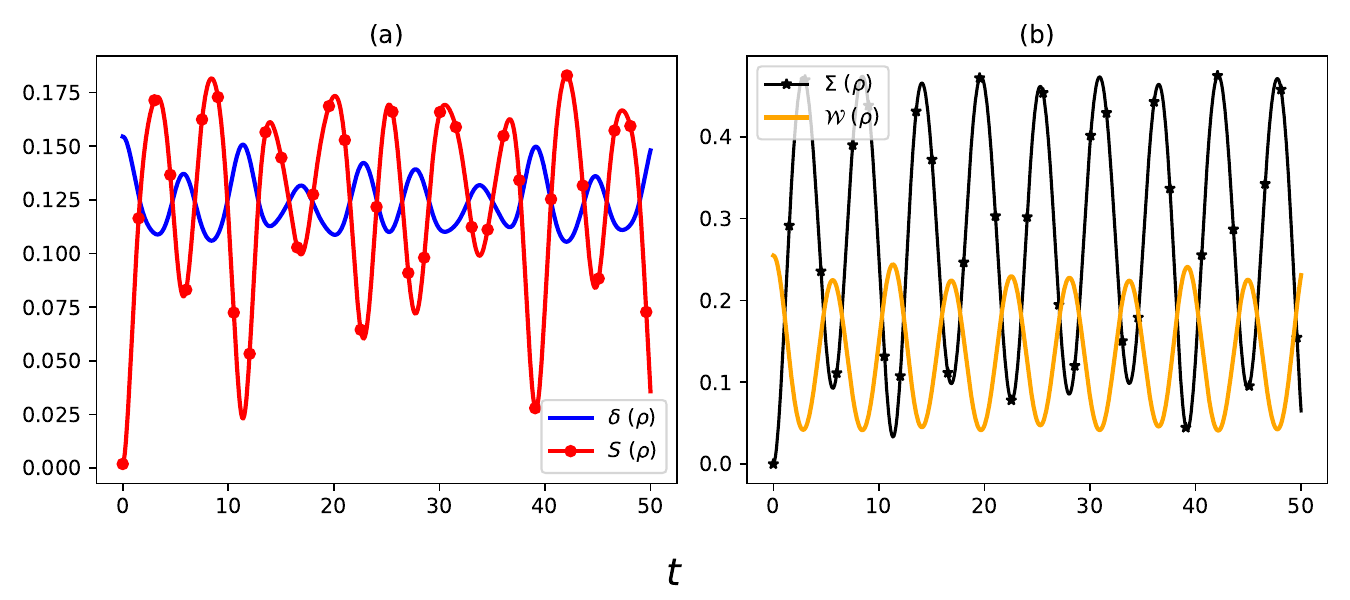}
     \caption{Variation of non-classical volume ($\delta$), von-Neumann entropy ($S$) in subplot $(a)$, and entropy production ($\Sigma$), ergotropy ($\mathcal{W}$) in subplot $(b)$ with time of the one-qubit maximally negative quantum state using the Jaynes-Cummings model. The parameters are: $\omega_0 = 1.5$, $\omega_c = 1$, $\beta = 3$, $g = 0.5$, and $\tau = 0.5$.}
    \label{Jaynes_cummings_model}
\end{figure}

The dynamics of $\delta$, $S$, $\Sigma$, and $\mathcal{W}$ for the single-qubit $NS_1$ state evolved using the Jaynes–Cummings Hamiltonian are analyzed using Eqs. \eqref{NV_formula}, \eqref{von_neumann_entropy}, \eqref{entropy_production_final}, and \eqref{ergotropy_formula}. As shown in Fig.~\ref{Jaynes_cummings_model}(a) and (b), this model also exhibits a consistent contradictory relationship, viz., $\delta$ evolves in opposition to $S$, while $\Sigma$ displays contradictory behavior with respect to $\mathcal{W}$. Moreover, the rise and fall patterns of $\delta$ and $\mathcal{W}$ and those of $S$ and $\Sigma$ remain in agreement with the trends observed in the collision model, central spin model, and the GAD channel. This consistency arises because the bosonic reservoir is initialized in a thermal state in the Jaynes-Cummings model as well.

\section{\label{conclusion}Conclusions}
In this article, the comparative analysis of different open quantum system models, namely the collision model, the central spin model, the spin–boson interaction model under the non-Markovian amplitude damping (NMAD) channel, the Markovian generalized amplitude damping (GAD) channel, and the Jaynes–Cummings model revealed a remarkable consistency in the thermodynamic interplay among non-classical volume $\delta$, von Neumann entropy $S$, entropy production $\Sigma$, and ergotropy $\mathcal{W}$. 
In all the cases, $\delta$ evolved in opposition to $S$. At the same time, $\Sigma$ displayed a contradictory relation with $\mathcal{W}$, emphasizing that enhanced randomness corresponds to reduced non-classicality and loss in accessible information and vice-versa. In contrast, reduced irreversibility allows for more work to be extracted.

Despite this universal structure, notable distinctions emerged depending on the nature of the reservoir. In the NMAD channel, where the bosonic bath was restricted to its ground state, correlated fluctuations between $\delta$ and $\mathcal{W}$, and $S$ and $\Sigma$ lose synchronization. In contrast, the collision model, central spin model, GAD channel, and Jaynes–Cummings model involving thermal reservoirs exhibited sustained, correlated fluctuations, preserving the interdependence among the four quantities.

These results highlight a universal interrelation among fundamental quantum thermodynamic and information quantifiers across different system environment interaction models. The non-classical volume $\delta$ evolves in opposition to von Neumann entropy $S$, while entropy production $\Sigma$ contrasts with ergotropy $\mathcal{W}$. The degree of correlation between entropy and entropy production, as well as between non-classical volume and ergotropy, also depends on the bath state, highlighting the crucial role of reservoir preparation in determining these quantifiers.


\bibliography{BibTexfile}
\bibliographystyle{apsrev4-2}

\end{document}